\documentclass[aps,prl,twocolumn,superscriptaddress,bibliography]{revtex4-2}
\usepackage{amsfonts}
\usepackage{mathrsfs}
\usepackage{amsmath}
\usepackage{xcolor}
\usepackage{graphicx}
\usepackage{bm}
\usepackage{amssymb}
\usepackage{xspace}
\usepackage{epstopdf}
\usepackage{dcolumn}
\usepackage{longtable}
\usepackage{multirow}
\usepackage{float}
\usepackage{comment}
\usepackage{wasysym}
\usepackage{esint}
\usepackage[colorlinks=true, letterpaper=true, pdfstartview=FitV,  linkcolor=blue, citecolor=blue, urlcolor=blue]{hyperref}

\makeatletter

\providecommand{\tabularnewline}{\\}

\makeatother

\usepackage{babel}
\begin{document}
\title{Theory of Nonlocal Transport from Nonlinear Valley Responses}

\author{Jin Cao}
\affiliation{Institute of Applied Physics and Materials Engineering, Faculty of Science and Technology, University of Macau, Macau SAR, China}

\author{Hui Wang}
\affiliation{Division of Physics and Applied Physics, School of Physical and Mathematical Sciences, Nanyang Technological University, Singapore 637371, Singapore}

\author{Shen Lai}
\email{laishen@um.edu.mo}
\affiliation{Institute of Applied Physics and Materials Engineering, Faculty of Science and Technology, University of Macau, Macau SAR, China}

\author{Cong Xiao}
\email{congxiao@fudan.edu.cn}
\affiliation{Interdisciplinary Center for Theoretical Physics and Information Sciences (ICTPIS), Fudan University, Shanghai 200433, China}

\author{Shengyuan A. Yang}
\affiliation{Institute of Applied Physics and Materials Engineering, Faculty of Science and Technology, University of Macau, Macau SAR, China}

\begin{abstract}
We develop a theory for the nonlocal measurement of nonlinear valley Hall effect.
Different from the linear case where the direct and the inverse processes are reciprocal, we unveil that the nonlinear inverse valley Hall effect needed to generate nonlocal voltage signal must have a distinct symmetry character and involve distinct mechanisms compared to the nonlinear valley Hall response it probes.
Particularly, it must be valley-even, in contrast to both linear and nonlinear valley Hall effects which are valley-odd.
Layer groups that permit such nonlocal valley responses are obtained via symmetry analysis, and formulas for the nonlocal signals are derived. In the presence of both linear and nonlinear valley responses, we show that
the different responses can be distinguished by their distinct scaling behaviors in the different harmonic components, under a low-frequency \emph{ac} driving. Combined with first-principles calculations, we predict sizable nonlocal transport signals from nonlinear valley responses in bilayer $T_{d}$-WTe$_{2}$. Our work lays a foundation for nonlocal transport studies on the emerging nonlinear valleytronics.
\end{abstract}

\maketitle
Valleytronics aims to exploit valley degree of freedom, referring to separated energy-degenerate portions in band structure, for new physical phenomena and applications~\cite{Rycerz2007Valley,Gunawan2006Valley,Xiao2007Valley,Yao2008Valley,Xiao2012Coupled,Zhu2011Field,Cai2013Magnetic,Xu2014Spin,Schaibley2016Valleytronics,Mak2018Light}. A key effect in valleytronics is valley Hall effect (VHE),
where a charge current generates a charge-neutral valley current flow in the transverse direction~\cite{Xiao2007Valley,Xiao2012Coupled,Mak2014valley,Lee2016Electrical}.
To probe this valley current by electric means, a common way is through the nonlocal measurement via an inverse VHE~\cite{Gorbachev2014Detecting,Sui2015Gate,Shimazaki2015Generation,Wu2019Intrinsic}.

\begin{figure}[b]
\begin{centering}
\includegraphics[width=8.6cm]{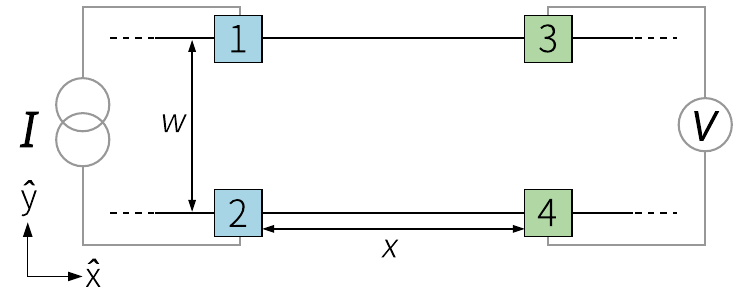}
\par\end{centering}
\caption{\label{fig1}Schematic of nonlocal measurement setup. The sample strip has a width $w$. A current $I$ is applied between contacts 1 and 2, and the voltage signal between 3 and 4 is measured.
}
\end{figure}

In a typical nonlocal transport setup,
as illustrated in Fig.~\ref{fig1}, a charge current is applied between contacts 1 and 2. Then, the valley current $j^v$ induced by VHE flows in $x$ direction and creates an imbalance of valley populations along the sample strip. By the inverse VHE, this valley imbalance is converted back to a voltage signal that is measured between 3 and 4 away from the applied-current region. As an essential component of this measurement, the inverse VHE shares the \emph{same} nature as VHE, i.e., they originate from the same microscopic mechanism and possess the same symmetry character. Indeed, as linear response effects, the two must have equal response coefficients, as dictated by Onsager's reciprocity principle.

Recently, there is a surge of interest in studying various nonlinear transport effects~\cite{Gao2014Field,Sodemann2015Quantum,Ma2018Observation,Kang2019Nonlinear,Du2021Nonlinear}. Nonlinear VHE has also been proposed \cite{Yu2014Nonlinear,Rodin2016Valley,Das2024Nonlinear}, which produces $j^v\propto E^2$. In crystals where linear VHE (hence also linear inverse VHE) is \emph{forbidden} by certain symmetry, the nonlinear VHE may become dominant. Then, a natural question is: Can such nonlinear response also be measured with the nonlocal setup?
Clearly, the crucial part is the inverse process. What will be the nonlinear version of inverse VHE that can produce the nonlocal voltage signal? Does it also share the same nature as the nonlinear VHE, as one might naively expect? These fundamental questions have not been addressed so far.

In this work, we answer these questions and develop the theory for nonlocal measurement of nonlinear VHE. We show that unlike the linear case, the nonlinear inverse VHE needed for nonlocal measurement must have a nature distinct from the
nonlinear VHE it probes. The nonlinear VHE and nonlinear inverse VHE must have distinct symmetry characters and originate from distinct mechanisms.
Specifically, the direct process is valley-odd, whereas the inverse process is valley-even. Based on this understanding, we derive a formula for the nonlocal signal and clarify its scaling with applied current and other system parameters. We also search through all layer groups (LG) and identify those permitting the nonlocal probe of nonlinear VHE.
In addition, we show that in systems where both linear and nonlinear VHEs are present, nonlinear signals can still be distinguished by low-frequency \emph{ac} modulation and by their unique scaling behaviors. Combining our theory with first-principles calculations, we estimate the nonlocal voltage signal induced by nonlinear VHE in bilayer WTe$_2$, obtaining a sizable result that can be detected in experiment.
Our work establishes the fundamental framework for nonlocal transport from nonlinear valley responses and opens up new opportunities for valleytronic applications.

\textit{\color{blue} Character of nonlinear inverse VHE.}
First, let us briefly review the picture of nonlocal measurement of linear VHE~\cite{Abanin2009Nonlocal,Beconcini2016Nonlocal}, focusing on the symmetry character of the physical processes involved.

Assume a metal or doped semiconductor (so that conduction is dominated by bulk carriers) having two valleys $A$ and $B$, connected by some symmetry. The VHE, which occurs in the region near contacts 1 and 2 in Fig.~\ref{fig1}, is described by
$
  j^v_x=\sigma^v_{H}E_y
$,
where $\sigma^v_H$ is the valley Hall conductivity,  $j^v_x=j^A_x-j^B_x$ with $j^\xi$ ($\xi=A, B$) the valley-resolved current density. One can easily see that $\sigma^v_{H}$ is valley-odd, namely, it is an odd function upon switching the valleys, since $j^v$ is valley-odd whereas $E$ field is valley-even.

The flow of this valley current along the strip leads to a spatial gradient in the valley population imbalance $n^v=n^A-n^B$, with $n^\xi$ as the carrier density from $\xi$ valley. The inverse VHE produces a charge current $j$ described by
$j_y=\sigma^{v'}_H F_x$, where $F_x=-\partial_x n^v/2\nu$ denotes the driving force of this process and $\nu$ is the density of states at Fermi level for a single valley. In this relation, $\sigma^{v'}_H$ is again valley-odd, because $j$ is valley-even while $F$ is valley-odd. As mentioned, according to Onsager's principle, the inverse process is actually described by the same
coefficient $\sigma^{v'}_H=\sigma^v_H$.
The resulting charge current then generates a voltage signal detected between contacts 3 and 4, away from 1 and 2.

Now, consider the case when the linear VHE (and hence its inverse) is forbidden by some symmetry of the system.
Then, the nonlinear VHE may dominate the valley response in the applied-current region, with
\begin{equation}\label{nv}
  j^v_x=\chi^v E_y^2.
\end{equation}
Again, this flow leads to finite gradient $\partial_x n^v$ downstream. The leading-order inverse process that converts it to a charge signal would be described by
\begin{equation}\label{nr}
  j_y=\zeta^v F_x^2.
\end{equation}

Comparing these two relations and with their linear counterparts, one observes a crucial difference. While $\chi^v$ for the nonlinear VHE, like $\sigma^v_H$, is valley-odd; $\zeta^v$ for the nonlinear inverse VHE has to be valley-even. This change originates from the change of driving factor in Eq.~(\ref{nr}): square of the valley-odd quantity $F$ makes a valley-even driving force $F^2$ for the nonlinear inverse process. Thus, $\chi^v$ and $\zeta^v$ must have distinct symmetry characters. It follows that
nonlinear VHE and the inverse VHE involved here must be \emph{distinct} and \emph{independent} processes, involving fundamentally different physical mechanisms.

Several remarks are in order here. First, the analysis above is completely general, based only on the symmetry character upon the switch of valleys. It is independent of detailed microscopic mechanisms, material systems, and the particular symmetry that connects the valleys.

Second, because $E^2$ has the same symmetry properties as $F^2$ in Eq.~(\ref{nr}),
the valley-even response $\zeta^v$ is in fact related to the nonlinear charge Hall conductivity \cite{Gao2014Field,Sodemann2015Quantum,Du2021Nonlinear}. By using the Einstein relation, one can show that the two equal when the Fermi surface consists only of the two valleys.
Notably, $\zeta^v$ requires broken inversion symmetry $\mathcal{P}$ of crystals. This property differs from the nonlinear VHE in Eq.~(\ref{nv}), which puts no requirement on inversion symmetry~\cite{Das2024Nonlinear}. It follows that
nonlinear VHE in centrosymmetric crystals cannot generate a nonlocal voltage signal via nonlinear inverse VHE.

Third, previous studies of valleytronics were mostly on systems where the two valleys are connected by time-reversal ($\mathcal T$) symmetry. In such a case, for a single valley, $\chi^v$ and $\zeta^v$ correspond to $\mathcal T$-odd and $\mathcal T$-even nonlinear charge Hall responses, respectively. These two types of nonlinear charge transport are a focus of recent research \cite{Gao2014Field,Sodemann2015Quantum,Ma2018Observation,Kang2019Nonlinear,Gao2023QM,Wang2023QM,Han2024room}, and are known to involve distinct mechanisms. For example, the $\mathcal T$-odd nonlinear response may contain an intrinsic contribution \cite{Gao2014Field,wang2021,liu2021}, which is independent of scattering and which does not exist in $\mathcal T$-even response. Extrinsic contributions for the two are also different and obey different scaling relations \cite{Du2019Disorder,Huang2023Scaling}.

\textit{\color{blue} Nonlocal transport signal.} Next, we derive a formula for the nonlocal voltage signal generated by the nonlinear VHE and inverse VHE discussed above.

Consider the nonlocal setup in Fig.~\ref{fig1} with a long strip sample. We work in the diffusive regime, where the sample length scale is much larger than the mean free path $l$. A charge current $I$ is applied between contacts 1 and 2, located at $x=0$.
It induces an inhomogeneous electric potential distribution $\phi(x,y)$ in the nearby region, which can be solved from the 2D Laplace equation $\nabla^2\phi=0$ (assuming charge neutrality condition) subjected to the boundary condition
$j_{y}\left(x,y=\pm w/2\right)=I\delta\left(x\right)$, where $w$ is the width of the strip. The solution is given by~\cite{Abanin2009Nonlocal}
\begin{equation}
\phi\left(x,y\right)=-I\int dk\,e^{ikx}\frac{\sinh\left(ky\right)}{2\pi\sigma k\cosh\left(kw/2\right)},\label{phi}
\end{equation}
where $\sigma$ is the local charge conductivity. Equation (\ref{phi}) gives the well-known nonlocal Ohmic (van der Pauw) resistance, which decays as $e^{-\pi x/w}$~\cite{Ihn2009Semiconductor}. To minimize its influence, one usually focuses on the region with $x\gg w$, which is assumed in the following.

On the other hand, the valley current generated by nonlinear VHE
can diffuse over a much longer distance, on the scale of valley diffusion length $\ell_{v}=\sqrt{D\tau_v}$ (assumed to be large compared to $w$), where $D$ is the diffusion constant and $\tau_v$ is the valley relaxation time.
The valley population imbalance can be solved from the diffusion equation
\begin{equation}
D\nabla^{2}{n}^{v}-{n}^{v}/\tau_{v}=\Xi,\label{DDE}
\end{equation}
where the source term (repeated indices are summed over)
\begin{equation}
\Xi\left(x,y\right)=\nabla\cdot j^v=\partial_{a}\left(\chi_{abc}^{v}E_{b}E_{c}\right)\label{rate}
\end{equation}
is resulted from the nonlinear valley current. Since we are interested in the voltage drop $\delta\phi$ at $x\gg w$,
it is convenient to integrate out the $y$ coordinate and convert Eq.~(\ref{DDE}) to a quasi-1D problem (details in Supplemental Material \cite{supp}).
When performing this integration for the source term, one may utilize the symmetry of the $E$ field distribution
[which corresponds to $\phi$ in Eq.~(\ref{phi})], namely, it resembles that of an electric dipole in the $y$ direction.
After integration, it is easy to see that only terms with $\chi_{xyy}^v$ and $\chi_{xxx}^v$ survive.

Solving the resulting diffusion equation, we obtain the valley imbalance distribution along the strip
\begin{equation}
n^{v}\left(x\right)\approx \frac{\tau_{v}I^{2}}{\pi w\sigma^{2}\ell_{v}^{2}}\tilde\chi^v e^{-x/\ell_{v}}.
\end{equation}
where $\tilde\chi^v=\chi_{xxx}^{v}+\chi_{xyy}^{v}$. Interestingly, besides $\chi_{xyy}^{v}$,
the longitudinal response $\chi_{xxx}^{v}$ also makes a contribution here. This is different from the linear case, where only the Hall response contributes~\cite{Abanin2009Nonlocal,Beconcini2016Nonlocal}.

Then, the voltage signal between contacts 3 and 4 can be obtained from $\delta\phi=j_y w/\sigma$, with $j_y$ from Eq.~(\ref{nr}). The final result can be expressed as
\begin{equation}
\delta\phi\left(x\right)\approx\frac{I^{4}\rho^{7}}{\pi^{2}w\ell_{v}^{2}}(\tilde\chi^v)^{2}\zeta^v e^{-2x/\ell_{v}},\label{NL1}
\end{equation}
where $\rho=1/\sigma$ is the local resistivity.

This result shows the following features. First, one observes that the response coefficients for direct and inverse processes do not enter the result in a symmetric manner, i.e., $\tilde\chi^v$ is squared but $\zeta^v$ is not. As a result, the
sign of $\delta\phi$ is determined by that of $\zeta^v$ for the nonlinear inverse VHE. This differs from the linear case, where~\cite{Abanin2009Nonlocal}
\begin{equation}\label{lin}
\delta\phi(x)\approx
(w/2\ell_v)I\rho^3(\sigma^v_H)^2 e^{-x/\ell_v}
\end{equation} having a fixed sign. Second, the scaling behavior here also differs from the linear response, e.g., the nonlinear processes here give $\delta\phi\propto I^4$, whereas the linear case $\propto I$. Finally, compared to the linear case, the exponent of the decay factor $e^{-2x/\ell_{v}}$ has an extra factor of 2, which stems from the nonlinearity of the inverse VHE.

\begin{table}[t]
\caption{\label{tabi} Symmetry constraints on linear and nonlinear valley conductivity tensors.  The prime indicates that the operation is of valley-switch type. $\checked$ ($\times$) means the response is symmetry-allowed (-forbidden).}
\begin{ruledtabular}
\begin{centering}
\begin{tabular}{ccccccc}
 & $\mathcal{P}^{\prime},\mathcal{C}_{2z}^{\prime}$ & $\mathcal{C}_{3z},\mathcal{M}_{z}$ & $\mathcal{C}_{2x}$ & $\mathcal{M}_{x}$ & $\mathcal{C}_{2x}^{\prime}$ & $\mathcal{M}_{x}^{\prime}$\tabularnewline
\hline
$\sigma^{v}_{H}$ & $\times$ & $\checked$ & $\times$ & $\times$ & $\checked$ & $\checked$\tabularnewline
$\chi_{xxx}^{v}$, $\chi_{xyy}^{v}$ & $\checked$ & $\checked$ & $\checked$ & $\times$ & $\times$ & $\checked$\tabularnewline
$\chi_{yxx}^{v}$, $\chi_{yyy}^{v}$ & $\checked$ & $\checked$ & $\times$ & $\checked$ & $\checked$ & $\times$\tabularnewline
$\zeta^v$ & $\times$ & $\checked$ & $\checked$ & $\checked$ & $\checked$ & $\checked$\tabularnewline
\end{tabular}
\par\end{centering}
\end{ruledtabular}
\end{table}

\textit{\color{blue}Symmetry constraint.}
Crystalline symmetries impose constraints on physical responses. Surprisingly, we note that even
the symmetry constraints for linear VHE have not been fully clarified yet.

Here, we focus on $\mathcal T$-invariant 2D systems.
We shall require that the system permits a valley structure with two and only two valleys connected by $\mathcal T$ operation.

To analyze valley-odd responses $\sigma^v_H$ and $\chi^v$, it is crucial to classify symmetries into two types: valley-preserve type and valley-switch type. As their names suggest, a valley-preserve (-switch) symmetry preserves (switches) the two valleys.
For example, inversion $\mathcal P$ and $\mathcal{C}_{2z}$ are valley-switch (we assume $z$ is normal to the 2D system), whereas $\mathcal{M}_z$ must be valley-preserve.

The transformation rule for $\chi^{v}$ is given by
\begin{eqnarray}
\chi_{abc}^{v} & = & \epsilon_v\mathcal{O}_{aa^{\prime}}\mathcal{O}_{bb^{\prime}}\mathcal{O}_{cc^{\prime}}\chi_{a^{\prime}b^{\prime}c^{\prime}}^{v},
\end{eqnarray}
where $\mathcal{O}$ is a point group operation, and $\epsilon_v=\pm$ if $\mathcal{O}$ is of valley-preserve/switch type.
The equation for $\sigma^v_H$ has a similar form. As for $\zeta^v$, since it is valley-even, we do not need to distinguish the valley type of $\mathcal{O}$.

In Table~\ref{tabi}, we present the constraints on the response coefficients from typical point group operations for 2D.
To distinguish valley-switch operations, we add a prime to the symmetry symbol. Some operations, e.g., $\mathcal C_{2x}$, does not have a fixed type. It may or may not preserve valleys, depending on the valley location. In general, the $x$ and $y$ directions for the strip geometry in Fig.~\ref{fig1} may differ from the crystal axis. We find that by a rotation, $\zeta^v_{yxx}$ can always be made nonzero as along as $\zeta^v_{abc}$ has any nonzero tensor component. The only symmetries that forbid $\zeta^v$ are
$\mathcal P$ and $\mathcal C_{2z}$. As for $\tilde\chi^v$ in (\ref{NL1}), the relevant tensor components are $\chi_{abb}^{v}$'s ($a,b=x,y$). Table~\ref{tabi} offers useful guidance for analyzing the symmetry character of response for specific materials.

On the basis of Table~\ref{tabi}, we screen through the 80 LGs. We find that in all LGs with broken $\mathcal{P}$ and $\mathcal{C}_{2z}$,  $\tilde\chi^v$ and $\zeta^v$ are both allowed. These LGs can be categorized into oblique lattices (LG~1, 4, and 5), rectangular lattices (LG~8-13 and 27-36), and hexagonal lattices (LG~65, 67-70, 74, 78, and 79). Among these candidates, linear VHE is forbidden in LG~68, 70, and 79. Linear VHE is also prohibited in those rectangular-lattice LGs, if rotation or mirror symmetry there is valley-preserve. The detailed discussion on the candidate LGs and the response tensors are presented in the Supplemental Material \cite{supp}.

\textit{\color{blue}Interplay with linear VHE.}
From the above symmetry analysis, we see that a number of crystalline systems allow both linear and nonlinear (inverse) VHEs.
Recent experiments on nonlinear transport demonstrate the possibility to distinguish nonlinear signals by using low-frequency \emph{ac} driving and lock-in technique \cite{Ma2018Observation,Kang2019Nonlinear,Lai2021}. Therefore, below, we investigate the nonlocal signal for a general scenario where both linear and nonlinear valley responses are present.

In such case, a low-frequency driving current $\tilde{I}=I\cos\omega t$ is applied between contacts 1 and 2, and the different
harmonic components of the nonlocal voltage signal
$\delta\tilde{\phi}=\sum_{n}\delta\phi_{n\omega}\cos (n\omega t)$ between 3 and 4 can be extracted using the lock-in technique. Via similar derivations which lead to (\ref{NL1}), we obtain the following results for the harmonic amplitudes
(retaining only the leading order terms in $I$)
\begin{eqnarray}
\delta\phi_{0/2\omega} & = & \frac{I^{2}\rho^{4}}{2\pi\ell_{v}}\tilde\chi^{v}\sigma^{v}_H e^{-x/\ell_{v}}+\frac{wI^{2}\rho^{5}}{8\ell_{v}^{2}}(\sigma^{v}_H)^{2}\zeta^v e^{-2x/\ell_{v}},\nonumber \\
\delta\phi_{3\omega} & = & \frac{I^{3}\rho^{6}}{4\pi\ell_{v}^{2}}\sigma^{v}_H \tilde\chi^{v}\zeta^v e^{-2x/\ell_{v}},\nonumber\\
\delta\phi_{4\omega} & = & \frac{I^{4}\rho^{7}}{8\pi^{2}w\ell_{v}^{2}}\left(\tilde\chi^{v}\right)^{2}\zeta^v e^{-2x/\ell_{v}},\label{ac}
\end{eqnarray}
and $\delta\phi_\omega$ has the same expression as Eq. (\ref{lin}).

One observes that the rectified component and the second harmonic component have the same amplitude. The two terms in $\delta\phi_{0/2\omega}$ have different decay lengths ($\ell_v$ vs $\ell_v/2$) and originate from two different processes. The first is from nonlinear VHE and linear inverse VHE, while the second is from linear VHE and nonlinear inverse VHE. The nonlinear direct and inverse VHEs also simultaneously contribute to the third and fourth harmonics. The $4\omega$ term has been explained above. As for the $3\omega$ signal, it derives from the sum frequency component combining both linear and nonlinear VHEs.

\begin{figure}[t]
\begin{centering}
\includegraphics[width=8.8cm]{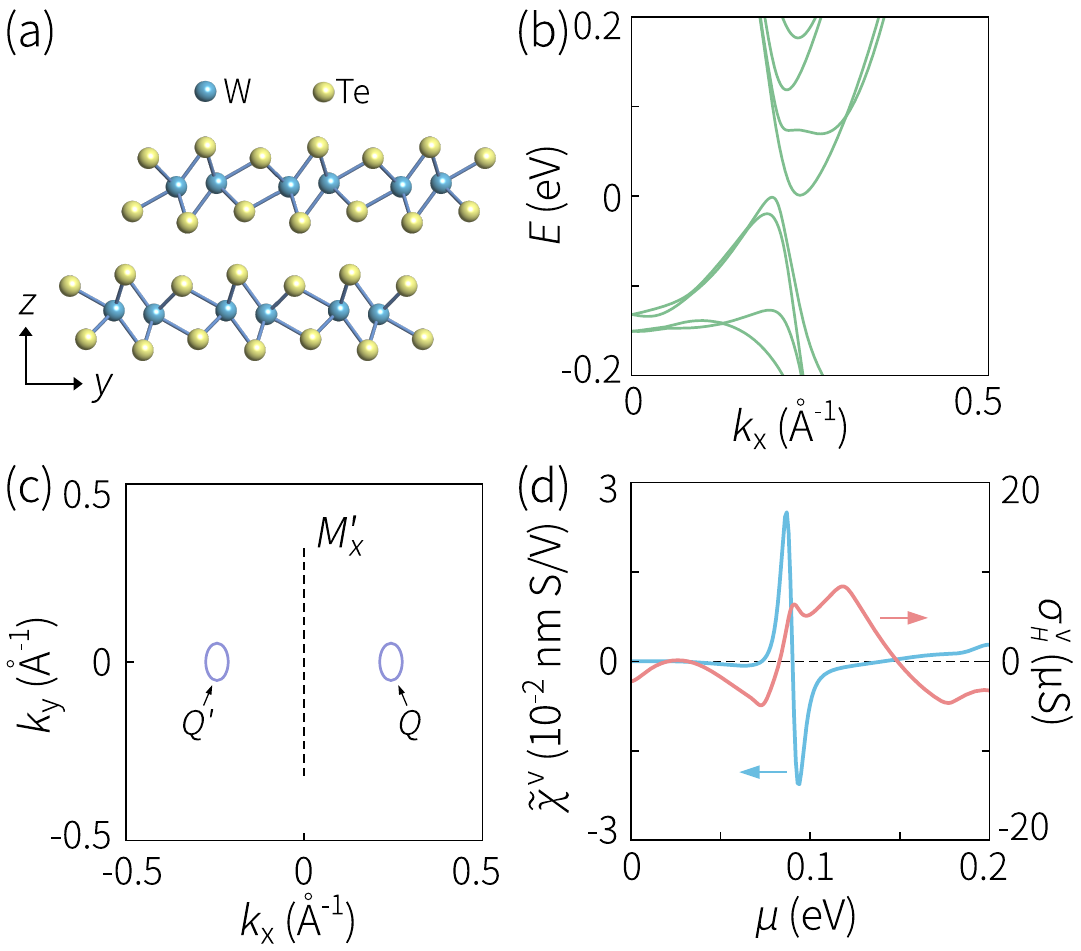}
\par\end{centering}
\caption{\label{fig2}(a) Crystal structure and (b) calculated low-energy band structure of bilayer $T_{d}$-WTe$_{2}$. (c) The Fermi surface at chemical potential of 40~meV, corresponding to the two valleys $Q$ and $Q^{\prime}$ connected by mirror symmetry. (d) Calculated intrinsic linear and nonlinear valley Hall conductivities versus chemical potential $\mu$.}
\end{figure}

\textit{\color{blue}First-principles evaluation for bilayer $\mathrm{WTe_{2}}$.} We perform an estimation of the nonlocal transport signal arising from nonlinear VHE. As noted from Eq.~(\ref{ac}), the nonlinear VHE is indispensable for $3\omega$ and $4\omega$ signals.
To have these signals, we also need a nonzero $\zeta^v$. As discussed above, this requires broken inversion symmetry, and for a system where the Fermi surface consists of only the two valleys, $\zeta^v$ has the same value as the nonlinear charge Hall conductivity.

From these considerations, we choose the example of bilayer WTe$_{2}$, whose nonlinear charge Hall conductivity has been measured in experiment \cite{Ma2018Observation}. Bilayer WTe$_{2}$ with $T_{d}$ stacking breaks inversion symmetry (Fig.~\ref{fig2}(a)). It belongs to  LG~11, which consists of a mirror symmetry $\mathcal{M}_{x}$. The low-energy band structure obtained from first-principles calculations is plotted in Fig.~\ref{fig2}(b) (calculation details in Supplemental Material \cite{supp}). It shows a very small indirect band gap $\sim 2$~meV around two valleys $Q$ and $Q'$ on the $\Gamma$-$X$ path (Fig.~\ref{fig2}(c)), consistent with previous studies \cite{Zheng2016Quantum,Ma2018Observation}.
As shown in Fig.~\ref{fig2}(c), $\mathcal{M}_{x}$ here is a valley-switch symmetry. According to Table~\ref{tabi}, both linear and nonlinear VHE responses are allowed. We shall focus on the $3\omega$ and $4\omega$ components and estimate their magnitudes from
Eq. (\ref{ac}).

We take a slight $n$ doping of the system with a chemical potential $\mu=0.09$ eV, which can be readily achieved by electric gating. We consider the intrinsic contribution to linear and nonlinear valley conductivities~\cite{Xiao2007Valley,Gao2014Field,wang2021,liu2021,Das2024Nonlinear}, and our calculation finds $\sigma^v_H\sim 6\ \mu$S and $\tilde\chi^v\sim 0.02$ nm$\cdot$S/V (see Fig.~\ref{fig2}(d)). As for $\zeta^v$, we take the experimental value of nonlinear charge Hall conductivity measured in this regime, which is $\sim 2$ nm$\cdot$S/V, along with $\rho\sim 10^4\ \Omega$ from experiment \cite{Ma2018Observation}. For a strip with $w\sim 0.3\ \mu$m, the magnitudes of $\delta\phi_{3\omega}$ and
$\delta\phi_{4\omega}$ will respectively be about $31~\mu$V and $16~\mu$V under a driving current of $I\sim 30~\mu$A, when measured at a distance on the order of $\ell_v$ (typically a few microns).
Such sizable voltage signals can be readily detected in experiments.

\textit{\color{blue}Discussion.} We have developed the theory for nonlocal transport mediated by nonlinear valley responses.
Distinct from the linear case, for nonlocal measurement of nonlinear VHE, the involved nonlinear inverse VHE is a process fundamentally different from the nonlinear VHE it probes. This feature leads to distinct scaling laws for the nonlocal voltage signal and to additional symmetry conditions on the material systems that can host such responses.
The candidate layer groups obtained from our symmetry analysis offer guidance for selecting material platforms. And the revealed
scaling laws will be essential for designing and interpreting experiments.
These findings not only clarify basic questions and establish a fundamental framework for nonlinear valley physics, but also pave the way towards novel valleytronic device applications.

Techniques for nonlocal transport measurement have been well developed.
Our theory, particularly Eq.~(\ref{ac}), provides the basis for analyzing experimental data.
For example, by making multiple voltage probes at varying locations, one may extract the valley diffusion length $\ell_v$ from $\delta\phi_\omega$ signal, as was done in previous experiments on linear VHE \cite{Gorbachev2014Detecting,Sui2015Gate,Shimazaki2015Generation,Wu2019Intrinsic}. Then, this information can be used to separate the two contributions in $\delta\phi_{0/2\omega}$ and to compare with the decay length for $\delta\phi_{3\omega/4\omega}$.
The scaling with driving current $I$ can be directly verified for different harmonics, whereas the scaling behavior with $\rho$ needs more care.
In experiment, one often changes $\rho$ by varying temperature, which mainly originates from the change in momentum relaxation time $\tau$. However, this change in $\tau$ may also affect the nonlinear coefficients $\tilde\chi^v$ and $\zeta^v$ (depending on their dominant mechanisms), thereby influencing the scaling of $\delta\phi$ when plotted as a function of $\rho$ $(\propto 1/\tau)$.
For example, suppose $\sigma^v_H,\tilde\chi^v\propto\tau^0$ are dominated by intrinsic~\cite{Xiao2007Valley,Gao2014Field,Das2024Nonlinear} or zeroth-order extrinsic mechanisms~\cite{Huang2023Scaling}, and suppose $\zeta^v\propto\tau$ is dominated by extrinsic Berry curvature dipole contribution \cite{Sodemann2015Quantum}. Then, one would obtain $\delta\phi_{\omega}\propto\rho^3$, $\delta\phi_{0/2\omega}\propto\rho^4$, $\delta\phi_{3\omega}\propto\rho^5$, and $\delta\phi_{4\omega}\propto\rho^6$.
It is worth noting that the scaling of $\zeta^v$ can be separately determined from local nonlinear charge transport measurement.
Such scaling analysis will be very helpful for revealing the microscopic mechanisms underlying nonlinear valley responses.


\bibliographystyle{apsrev4-2}
\bibliography{ref.bib}

\begin{thebibliography}{37}%
\makeatletter
\providecommand \@ifxundefined [1]{%
 \@ifx{#1\undefined}
}%
\providecommand \@ifnum [1]{%
 \ifnum #1\expandafter \@firstoftwo
 \else \expandafter \@secondoftwo
 \fi
}%
\providecommand \@ifx [1]{%
 \ifx #1\expandafter \@firstoftwo
 \else \expandafter \@secondoftwo
 \fi
}%
\providecommand \natexlab [1]{#1}%
\providecommand \enquote  [1]{``#1''}%
\providecommand \bibnamefont  [1]{#1}%
\providecommand \bibfnamefont [1]{#1}%
\providecommand \citenamefont [1]{#1}%
\providecommand \href@noop [0]{\@secondoftwo}%
\providecommand \href [0]{\begingroup \@sanitize@url \@href}%
\providecommand \@href[1]{\@@startlink{#1}\@@href}%
\providecommand \@@href[1]{\endgroup#1\@@endlink}%
\providecommand \@sanitize@url [0]{\catcode `\\12\catcode `\$12\catcode
  `\&12\catcode `\#12\catcode `\^12\catcode `\_12\catcode `\%12\relax}%
\providecommand \@@startlink[1]{}%
\providecommand \@@endlink[0]{}%
\providecommand \url  [0]{\begingroup\@sanitize@url \@url }%
\providecommand \@url [1]{\endgroup\@href {#1}{\urlprefix }}%
\providecommand \urlprefix  [0]{URL }%
\providecommand \Eprint [0]{\href }%
\providecommand \doibase [0]{https://doi.org/}%
\providecommand \selectlanguage [0]{\@gobble}%
\providecommand \bibinfo  [0]{\@secondoftwo}%
\providecommand \bibfield  [0]{\@secondoftwo}%
\providecommand \translation [1]{[#1]}%
\providecommand \BibitemOpen [0]{}%
\providecommand \bibitemStop [0]{}%
\providecommand \bibitemNoStop [0]{.\EOS\space}%
\providecommand \EOS [0]{\spacefactor3000\relax}%
\providecommand \BibitemShut  [1]{\csname bibitem#1\endcsname}%
\let\auto@bib@innerbib\@empty
\bibitem [{\citenamefont {Rycerz}\ \emph {et~al.}(2007)\citenamefont {Rycerz},
  \citenamefont {Tworzyd{\l}o},\ and\ \citenamefont
  {Beenakker}}]{Rycerz2007Valley}%
  \BibitemOpen
  \bibfield  {author} {\bibinfo {author} {\bibfnamefont {A.}~\bibnamefont
  {Rycerz}}, \bibinfo {author} {\bibfnamefont {J.}~\bibnamefont
  {Tworzyd{\l}o}},\ and\ \bibinfo {author} {\bibfnamefont {C.~W.~J.}\
  \bibnamefont {Beenakker}},\ }\href {https://doi.org/10.1038/nphys547}
  {\bibfield  {journal} {\bibinfo  {journal} {Nature Physics}\ }\textbf
  {\bibinfo {volume} {3}},\ \bibinfo {pages} {172} (\bibinfo {year}
  {2007})}\BibitemShut {NoStop}%
\bibitem [{\citenamefont {Gunawan}\ \emph {et~al.}(2006)\citenamefont
  {Gunawan}, \citenamefont {Shkolnikov}, \citenamefont {Vakili}, \citenamefont
  {Gokmen}, \citenamefont {De~Poortere},\ and\ \citenamefont
  {Shayegan}}]{Gunawan2006Valley}%
  \BibitemOpen
  \bibfield  {author} {\bibinfo {author} {\bibfnamefont {O.}~\bibnamefont
  {Gunawan}}, \bibinfo {author} {\bibfnamefont {Y.~P.}\ \bibnamefont
  {Shkolnikov}}, \bibinfo {author} {\bibfnamefont {K.}~\bibnamefont {Vakili}},
  \bibinfo {author} {\bibfnamefont {T.}~\bibnamefont {Gokmen}}, \bibinfo
  {author} {\bibfnamefont {E.~P.}\ \bibnamefont {De~Poortere}},\ and\ \bibinfo
  {author} {\bibfnamefont {M.}~\bibnamefont {Shayegan}},\ }\href
  {https://doi.org/10.1103/PhysRevLett.97.186404} {\bibfield  {journal}
  {\bibinfo  {journal} {Phys. Rev. Lett.}\ }\textbf {\bibinfo {volume} {97}},\
  \bibinfo {pages} {186404} (\bibinfo {year} {2006})}\BibitemShut {NoStop}%
\bibitem [{\citenamefont {Xiao}\ \emph {et~al.}(2007)\citenamefont {Xiao},
  \citenamefont {Yao},\ and\ \citenamefont {Niu}}]{Xiao2007Valley}%
  \BibitemOpen
  \bibfield  {author} {\bibinfo {author} {\bibfnamefont {D.}~\bibnamefont
  {Xiao}}, \bibinfo {author} {\bibfnamefont {W.}~\bibnamefont {Yao}},\ and\
  \bibinfo {author} {\bibfnamefont {Q.}~\bibnamefont {Niu}},\ }\href
  {https://doi.org/10.1103/PhysRevLett.99.236809} {\bibfield  {journal}
  {\bibinfo  {journal} {Phys. Rev. Lett.}\ }\textbf {\bibinfo {volume} {99}},\
  \bibinfo {pages} {236809} (\bibinfo {year} {2007})}\BibitemShut {NoStop}%
\bibitem [{\citenamefont {Yao}\ \emph {et~al.}(2008)\citenamefont {Yao},
  \citenamefont {Xiao},\ and\ \citenamefont {Niu}}]{Yao2008Valley}%
  \BibitemOpen
  \bibfield  {author} {\bibinfo {author} {\bibfnamefont {W.}~\bibnamefont
  {Yao}}, \bibinfo {author} {\bibfnamefont {D.}~\bibnamefont {Xiao}},\ and\
  \bibinfo {author} {\bibfnamefont {Q.}~\bibnamefont {Niu}},\ }\href
  {https://doi.org/10.1103/PhysRevB.77.235406} {\bibfield  {journal} {\bibinfo
  {journal} {Phys. Rev. B}\ }\textbf {\bibinfo {volume} {77}},\ \bibinfo
  {pages} {235406} (\bibinfo {year} {2008})}\BibitemShut {NoStop}%
\bibitem [{\citenamefont {Xiao}\ \emph {et~al.}(2012)\citenamefont {Xiao},
  \citenamefont {Liu}, \citenamefont {Feng}, \citenamefont {Xu},\ and\
  \citenamefont {Yao}}]{Xiao2012Coupled}%
  \BibitemOpen
  \bibfield  {author} {\bibinfo {author} {\bibfnamefont {D.}~\bibnamefont
  {Xiao}}, \bibinfo {author} {\bibfnamefont {G.-B.}\ \bibnamefont {Liu}},
  \bibinfo {author} {\bibfnamefont {W.}~\bibnamefont {Feng}}, \bibinfo {author}
  {\bibfnamefont {X.}~\bibnamefont {Xu}},\ and\ \bibinfo {author}
  {\bibfnamefont {W.}~\bibnamefont {Yao}},\ }\href
  {https://doi.org/10.1103/PhysRevLett.108.196802} {\bibfield  {journal}
  {\bibinfo  {journal} {Phys. Rev. Lett.}\ }\textbf {\bibinfo {volume} {108}},\
  \bibinfo {pages} {196802} (\bibinfo {year} {2012})}\BibitemShut {NoStop}%
\bibitem [{\citenamefont {Zhu}\ \emph {et~al.}(2012)\citenamefont {Zhu},
  \citenamefont {Collaudin}, \citenamefont {Fauqué}, \citenamefont {Kang},\
  and\ \citenamefont {Behnia}}]{Zhu2011Field}%
  \BibitemOpen
  \bibfield  {author} {\bibinfo {author} {\bibfnamefont {Z.}~\bibnamefont
  {Zhu}}, \bibinfo {author} {\bibfnamefont {A.}~\bibnamefont {Collaudin}},
  \bibinfo {author} {\bibfnamefont {B.}~\bibnamefont {Fauqué}}, \bibinfo
  {author} {\bibfnamefont {W.}~\bibnamefont {Kang}},\ and\ \bibinfo {author}
  {\bibfnamefont {K.}~\bibnamefont {Behnia}},\ }\href
  {https://doi.org/10.1038/nphys2111} {\bibfield  {journal} {\bibinfo
  {journal} {Nature Physics}\ }\textbf {\bibinfo {volume} {8}},\ \bibinfo
  {pages} {89} (\bibinfo {year} {2012})}\BibitemShut {NoStop}%
\bibitem [{\citenamefont {Cai}\ \emph {et~al.}(2013)\citenamefont {Cai},
  \citenamefont {Yang}, \citenamefont {Li}, \citenamefont {Zhang},
  \citenamefont {Shi}, \citenamefont {Yao},\ and\ \citenamefont
  {Niu}}]{Cai2013Magnetic}%
  \BibitemOpen
  \bibfield  {author} {\bibinfo {author} {\bibfnamefont {T.}~\bibnamefont
  {Cai}}, \bibinfo {author} {\bibfnamefont {S.~A.}\ \bibnamefont {Yang}},
  \bibinfo {author} {\bibfnamefont {X.}~\bibnamefont {Li}}, \bibinfo {author}
  {\bibfnamefont {F.}~\bibnamefont {Zhang}}, \bibinfo {author} {\bibfnamefont
  {J.}~\bibnamefont {Shi}}, \bibinfo {author} {\bibfnamefont {W.}~\bibnamefont
  {Yao}},\ and\ \bibinfo {author} {\bibfnamefont {Q.}~\bibnamefont {Niu}},\
  }\href {https://doi.org/10.1103/PhysRevB.88.115140} {\bibfield  {journal}
  {\bibinfo  {journal} {Phys. Rev. B}\ }\textbf {\bibinfo {volume} {88}},\
  \bibinfo {pages} {115140} (\bibinfo {year} {2013})}\BibitemShut {NoStop}%
\bibitem [{\citenamefont {Xu}\ \emph {et~al.}(2014)\citenamefont {Xu},
  \citenamefont {Yao}, \citenamefont {Xiao},\ and\ \citenamefont
  {Heinz}}]{Xu2014Spin}%
  \BibitemOpen
  \bibfield  {author} {\bibinfo {author} {\bibfnamefont {X.}~\bibnamefont
  {Xu}}, \bibinfo {author} {\bibfnamefont {W.}~\bibnamefont {Yao}}, \bibinfo
  {author} {\bibfnamefont {D.}~\bibnamefont {Xiao}},\ and\ \bibinfo {author}
  {\bibfnamefont {T.~F.}\ \bibnamefont {Heinz}},\ }\href
  {https://doi.org/10.1038/nphys2942} {\bibfield  {journal} {\bibinfo
  {journal} {Nature Physics}\ }\textbf {\bibinfo {volume} {10}},\ \bibinfo
  {pages} {343} (\bibinfo {year} {2014})}\BibitemShut {NoStop}%
\bibitem [{\citenamefont {Schaibley}\ \emph {et~al.}(2016)\citenamefont
  {Schaibley}, \citenamefont {Yu}, \citenamefont {Clark}, \citenamefont
  {Rivera}, \citenamefont {Ross}, \citenamefont {Seyler}, \citenamefont {Yao},\
  and\ \citenamefont {Xu}}]{Schaibley2016Valleytronics}%
  \BibitemOpen
  \bibfield  {author} {\bibinfo {author} {\bibfnamefont {J.~R.}\ \bibnamefont
  {Schaibley}}, \bibinfo {author} {\bibfnamefont {H.}~\bibnamefont {Yu}},
  \bibinfo {author} {\bibfnamefont {G.}~\bibnamefont {Clark}}, \bibinfo
  {author} {\bibfnamefont {P.}~\bibnamefont {Rivera}}, \bibinfo {author}
  {\bibfnamefont {J.~S.}\ \bibnamefont {Ross}}, \bibinfo {author}
  {\bibfnamefont {K.~L.}\ \bibnamefont {Seyler}}, \bibinfo {author}
  {\bibfnamefont {W.}~\bibnamefont {Yao}},\ and\ \bibinfo {author}
  {\bibfnamefont {X.}~\bibnamefont {Xu}},\ }\href
  {https://doi.org/10.1038/natrevmats.2016.55} {\bibfield  {journal} {\bibinfo
  {journal} {Nature Reviews Materials}\ }\textbf {\bibinfo {volume} {1}},\
  \bibinfo {pages} {16055} (\bibinfo {year} {2016})}\BibitemShut {NoStop}%
\bibitem [{\citenamefont {Mak}\ \emph {et~al.}(2018)\citenamefont {Mak},
  \citenamefont {Xiao},\ and\ \citenamefont {Shan}}]{Mak2018Light}%
  \BibitemOpen
  \bibfield  {author} {\bibinfo {author} {\bibfnamefont {K.~F.}\ \bibnamefont
  {Mak}}, \bibinfo {author} {\bibfnamefont {D.}~\bibnamefont {Xiao}},\ and\
  \bibinfo {author} {\bibfnamefont {J.}~\bibnamefont {Shan}},\ }\href
  {https://doi.org/10.1038/s41566-018-0204-6} {\bibfield  {journal} {\bibinfo
  {journal} {Nature Photonics}\ }\textbf {\bibinfo {volume} {12}},\ \bibinfo
  {pages} {451} (\bibinfo {year} {2018})}\BibitemShut {NoStop}%
\bibitem [{\citenamefont {Mak}\ \emph {et~al.}(2014)\citenamefont {Mak},
  \citenamefont {McGill}, \citenamefont {Park},\ and\ \citenamefont
  {McEuen}}]{Mak2014valley}%
  \BibitemOpen
  \bibfield  {author} {\bibinfo {author} {\bibfnamefont {K.~F.}\ \bibnamefont
  {Mak}}, \bibinfo {author} {\bibfnamefont {K.~L.}\ \bibnamefont {McGill}},
  \bibinfo {author} {\bibfnamefont {J.}~\bibnamefont {Park}},\ and\ \bibinfo
  {author} {\bibfnamefont {P.~L.}\ \bibnamefont {McEuen}},\ }\href
  {https://doi.org/10.1126/science.1250140} {\bibfield  {journal} {\bibinfo
  {journal} {Science}\ }\textbf {\bibinfo {volume} {344}},\ \bibinfo {pages}
  {1489} (\bibinfo {year} {2014})}\BibitemShut {NoStop}%
\bibitem [{\citenamefont {Lee}\ \emph {et~al.}(2016)\citenamefont {Lee},
  \citenamefont {Mak},\ and\ \citenamefont {Shan}}]{Lee2016Electrical}%
  \BibitemOpen
  \bibfield  {author} {\bibinfo {author} {\bibfnamefont {J.}~\bibnamefont
  {Lee}}, \bibinfo {author} {\bibfnamefont {K.~F.}\ \bibnamefont {Mak}},\ and\
  \bibinfo {author} {\bibfnamefont {J.}~\bibnamefont {Shan}},\ }\href
  {https://doi.org/10.1038/nnano.2015.337} {\bibfield  {journal} {\bibinfo
  {journal} {Nature Nanotechnology}\ }\textbf {\bibinfo {volume} {11}},\
  \bibinfo {pages} {421} (\bibinfo {year} {2016})}\BibitemShut {NoStop}%
\bibitem [{\citenamefont {Gorbachev}\ \emph {et~al.}(2014)\citenamefont
  {Gorbachev}, \citenamefont {Song}, \citenamefont {Yu}, \citenamefont
  {Kretinin}, \citenamefont {Withers}, \citenamefont {Cao}, \citenamefont
  {Mishchenko}, \citenamefont {Grigorieva}, \citenamefont {Novoselov},
  \citenamefont {Levitov},\ and\ \citenamefont
  {Geim}}]{Gorbachev2014Detecting}%
  \BibitemOpen
  \bibfield  {author} {\bibinfo {author} {\bibfnamefont {R.~V.}\ \bibnamefont
  {Gorbachev}}, \bibinfo {author} {\bibfnamefont {J.~C.~W.}\ \bibnamefont
  {Song}}, \bibinfo {author} {\bibfnamefont {G.~L.}\ \bibnamefont {Yu}},
  \bibinfo {author} {\bibfnamefont {A.~V.}\ \bibnamefont {Kretinin}}, \bibinfo
  {author} {\bibfnamefont {F.}~\bibnamefont {Withers}}, \bibinfo {author}
  {\bibfnamefont {Y.}~\bibnamefont {Cao}}, \bibinfo {author} {\bibfnamefont
  {A.}~\bibnamefont {Mishchenko}}, \bibinfo {author} {\bibfnamefont {I.~V.}\
  \bibnamefont {Grigorieva}}, \bibinfo {author} {\bibfnamefont {K.~S.}\
  \bibnamefont {Novoselov}}, \bibinfo {author} {\bibfnamefont {L.~S.}\
  \bibnamefont {Levitov}},\ and\ \bibinfo {author} {\bibfnamefont {A.~K.}\
  \bibnamefont {Geim}},\ }\href {https://doi.org/10.1126/science.1254966}
  {\bibfield  {journal} {\bibinfo  {journal} {Science}\ }\textbf {\bibinfo
  {volume} {346}},\ \bibinfo {pages} {448} (\bibinfo {year}
  {2014})}\BibitemShut {NoStop}%
\bibitem [{\citenamefont {Sui}\ \emph {et~al.}(2015)\citenamefont {Sui},
  \citenamefont {Chen}, \citenamefont {Ma}, \citenamefont {Shan}, \citenamefont
  {Tian}, \citenamefont {Watanabe}, \citenamefont {Taniguchi}, \citenamefont
  {Jin}, \citenamefont {Yao}, \citenamefont {Xiao},\ and\ \citenamefont
  {Zhang}}]{Sui2015Gate}%
  \BibitemOpen
  \bibfield  {author} {\bibinfo {author} {\bibfnamefont {M.}~\bibnamefont
  {Sui}}, \bibinfo {author} {\bibfnamefont {G.}~\bibnamefont {Chen}}, \bibinfo
  {author} {\bibfnamefont {L.}~\bibnamefont {Ma}}, \bibinfo {author}
  {\bibfnamefont {W.-Y.}\ \bibnamefont {Shan}}, \bibinfo {author}
  {\bibfnamefont {D.}~\bibnamefont {Tian}}, \bibinfo {author} {\bibfnamefont
  {K.}~\bibnamefont {Watanabe}}, \bibinfo {author} {\bibfnamefont
  {T.}~\bibnamefont {Taniguchi}}, \bibinfo {author} {\bibfnamefont
  {X.}~\bibnamefont {Jin}}, \bibinfo {author} {\bibfnamefont {W.}~\bibnamefont
  {Yao}}, \bibinfo {author} {\bibfnamefont {D.}~\bibnamefont {Xiao}},\ and\
  \bibinfo {author} {\bibfnamefont {Y.}~\bibnamefont {Zhang}},\ }\href
  {https://doi.org/10.1038/nphys3485} {\bibfield  {journal} {\bibinfo
  {journal} {Nature Physics}\ }\textbf {\bibinfo {volume} {11}},\ \bibinfo
  {pages} {1027} (\bibinfo {year} {2015})}\BibitemShut {NoStop}%
\bibitem [{\citenamefont {Shimazaki}\ \emph {et~al.}(2015)\citenamefont
  {Shimazaki}, \citenamefont {Yamamoto}, \citenamefont {Borzenets},
  \citenamefont {Watanabe}, \citenamefont {Taniguchi},\ and\ \citenamefont
  {Tarucha}}]{Shimazaki2015Generation}%
  \BibitemOpen
  \bibfield  {author} {\bibinfo {author} {\bibfnamefont {Y.}~\bibnamefont
  {Shimazaki}}, \bibinfo {author} {\bibfnamefont {M.}~\bibnamefont {Yamamoto}},
  \bibinfo {author} {\bibfnamefont {I.~V.}\ \bibnamefont {Borzenets}}, \bibinfo
  {author} {\bibfnamefont {K.}~\bibnamefont {Watanabe}}, \bibinfo {author}
  {\bibfnamefont {T.}~\bibnamefont {Taniguchi}},\ and\ \bibinfo {author}
  {\bibfnamefont {S.}~\bibnamefont {Tarucha}},\ }\href
  {https://doi.org/10.1038/nphys3551} {\bibfield  {journal} {\bibinfo
  {journal} {Nature Physics}\ }\textbf {\bibinfo {volume} {11}},\ \bibinfo
  {pages} {1032} (\bibinfo {year} {2015})}\BibitemShut {NoStop}%
\bibitem [{\citenamefont {Wu}\ \emph {et~al.}(2019)\citenamefont {Wu},
  \citenamefont {Zhou}, \citenamefont {Cai}, \citenamefont {Cheung},
  \citenamefont {Liu}, \citenamefont {Huang}, \citenamefont {Lin},
  \citenamefont {Han}, \citenamefont {An}, \citenamefont {Wang}, \citenamefont
  {Xu}, \citenamefont {Long}, \citenamefont {Cheng}, \citenamefont {Law},
  \citenamefont {Zhang},\ and\ \citenamefont {Wang}}]{Wu2019Intrinsic}%
  \BibitemOpen
  \bibfield  {author} {\bibinfo {author} {\bibfnamefont {Z.}~\bibnamefont
  {Wu}}, \bibinfo {author} {\bibfnamefont {B.~T.}\ \bibnamefont {Zhou}},
  \bibinfo {author} {\bibfnamefont {X.}~\bibnamefont {Cai}}, \bibinfo {author}
  {\bibfnamefont {P.}~\bibnamefont {Cheung}}, \bibinfo {author} {\bibfnamefont
  {G.-B.}\ \bibnamefont {Liu}}, \bibinfo {author} {\bibfnamefont
  {M.}~\bibnamefont {Huang}}, \bibinfo {author} {\bibfnamefont
  {J.}~\bibnamefont {Lin}}, \bibinfo {author} {\bibfnamefont {T.}~\bibnamefont
  {Han}}, \bibinfo {author} {\bibfnamefont {L.}~\bibnamefont {An}}, \bibinfo
  {author} {\bibfnamefont {Y.}~\bibnamefont {Wang}}, \bibinfo {author}
  {\bibfnamefont {S.}~\bibnamefont {Xu}}, \bibinfo {author} {\bibfnamefont
  {G.}~\bibnamefont {Long}}, \bibinfo {author} {\bibfnamefont {C.}~\bibnamefont
  {Cheng}}, \bibinfo {author} {\bibfnamefont {K.~T.}\ \bibnamefont {Law}},
  \bibinfo {author} {\bibfnamefont {F.}~\bibnamefont {Zhang}},\ and\ \bibinfo
  {author} {\bibfnamefont {N.}~\bibnamefont {Wang}},\ }\href
  {https://doi.org/10.1038/s41467-019-08629-9} {\bibfield  {journal} {\bibinfo
  {journal} {Nature Communications}\ }\textbf {\bibinfo {volume} {10}},\
  \bibinfo {pages} {611} (\bibinfo {year} {2019})}\BibitemShut {NoStop}%
\bibitem [{\citenamefont {Gao}\ \emph {et~al.}(2014)\citenamefont {Gao},
  \citenamefont {Yang},\ and\ \citenamefont {Niu}}]{Gao2014Field}%
  \BibitemOpen
  \bibfield  {author} {\bibinfo {author} {\bibfnamefont {Y.}~\bibnamefont
  {Gao}}, \bibinfo {author} {\bibfnamefont {S.~A.}\ \bibnamefont {Yang}},\ and\
  \bibinfo {author} {\bibfnamefont {Q.}~\bibnamefont {Niu}},\ }\href
  {https://doi.org/10.1103/PhysRevLett.112.166601} {\bibfield  {journal}
  {\bibinfo  {journal} {Phys. Rev. Lett.}\ }\textbf {\bibinfo {volume} {112}},\
  \bibinfo {pages} {166601} (\bibinfo {year} {2014})}\BibitemShut {NoStop}%
\bibitem [{\citenamefont {Sodemann}\ and\ \citenamefont
  {Fu}(2015)}]{Sodemann2015Quantum}%
  \BibitemOpen
  \bibfield  {author} {\bibinfo {author} {\bibfnamefont {I.}~\bibnamefont
  {Sodemann}}\ and\ \bibinfo {author} {\bibfnamefont {L.}~\bibnamefont {Fu}},\
  }\href {https://doi.org/10.1103/PhysRevLett.115.216806} {\bibfield  {journal}
  {\bibinfo  {journal} {Phys. Rev. Lett.}\ }\textbf {\bibinfo {volume} {115}},\
  \bibinfo {pages} {216806} (\bibinfo {year} {2015})}\BibitemShut {NoStop}%
\bibitem [{\citenamefont {Ma}\ \emph {et~al.}(2018)\citenamefont {Ma},
  \citenamefont {Xu}, \citenamefont {Shen}, \citenamefont {MacNeill},
  \citenamefont {Fatemi}, \citenamefont {Chang}, \citenamefont {Mier~Valdivia},
  \citenamefont {Wu}, \citenamefont {Du}, \citenamefont {Hsu}, \citenamefont
  {Fang}, \citenamefont {Gibson}, \citenamefont {Watanabe}, \citenamefont
  {Taniguchi}, \citenamefont {Cava}, \citenamefont {Kaxiras}, \citenamefont
  {Lu}, \citenamefont {Lin}, \citenamefont {Fu}, \citenamefont {Gedik},\ and\
  \citenamefont {Jarillo-Herrero}}]{Ma2018Observation}%
  \BibitemOpen
  \bibfield  {author} {\bibinfo {author} {\bibfnamefont {Q.}~\bibnamefont
  {Ma}}, \bibinfo {author} {\bibfnamefont {S.-Y.}\ \bibnamefont {Xu}}, \bibinfo
  {author} {\bibfnamefont {H.}~\bibnamefont {Shen}}, \bibinfo {author}
  {\bibfnamefont {D.}~\bibnamefont {MacNeill}}, \bibinfo {author}
  {\bibfnamefont {V.}~\bibnamefont {Fatemi}}, \bibinfo {author} {\bibfnamefont
  {T.-R.}\ \bibnamefont {Chang}}, \bibinfo {author} {\bibfnamefont {A.~M.}\
  \bibnamefont {Mier~Valdivia}}, \bibinfo {author} {\bibfnamefont
  {S.}~\bibnamefont {Wu}}, \bibinfo {author} {\bibfnamefont {Z.}~\bibnamefont
  {Du}}, \bibinfo {author} {\bibfnamefont {C.-H.}\ \bibnamefont {Hsu}},
  \bibinfo {author} {\bibfnamefont {S.}~\bibnamefont {Fang}}, \bibinfo {author}
  {\bibfnamefont {Q.~D.}\ \bibnamefont {Gibson}}, \bibinfo {author}
  {\bibfnamefont {K.}~\bibnamefont {Watanabe}}, \bibinfo {author}
  {\bibfnamefont {T.}~\bibnamefont {Taniguchi}}, \bibinfo {author}
  {\bibfnamefont {R.~J.}\ \bibnamefont {Cava}}, \bibinfo {author}
  {\bibfnamefont {E.}~\bibnamefont {Kaxiras}}, \bibinfo {author} {\bibfnamefont
  {H.-Z.}\ \bibnamefont {Lu}}, \bibinfo {author} {\bibfnamefont
  {H.}~\bibnamefont {Lin}}, \bibinfo {author} {\bibfnamefont {L.}~\bibnamefont
  {Fu}}, \bibinfo {author} {\bibfnamefont {N.}~\bibnamefont {Gedik}},\ and\
  \bibinfo {author} {\bibfnamefont {P.}~\bibnamefont {Jarillo-Herrero}},\
  }\href {https://doi.org/10.1038/s41586-018-0807-6} {\bibfield  {journal}
  {\bibinfo  {journal} {Nature}\ }\textbf {\bibinfo {volume} {565}},\ \bibinfo
  {pages} {337} (\bibinfo {year} {2018})}\BibitemShut {NoStop}%
\bibitem [{\citenamefont {Kang}\ \emph {et~al.}(2019)\citenamefont {Kang},
  \citenamefont {Li}, \citenamefont {Sohn}, \citenamefont {Shan},\ and\
  \citenamefont {Mak}}]{Kang2019Nonlinear}%
  \BibitemOpen
  \bibfield  {author} {\bibinfo {author} {\bibfnamefont {K.}~\bibnamefont
  {Kang}}, \bibinfo {author} {\bibfnamefont {T.}~\bibnamefont {Li}}, \bibinfo
  {author} {\bibfnamefont {E.}~\bibnamefont {Sohn}}, \bibinfo {author}
  {\bibfnamefont {J.}~\bibnamefont {Shan}},\ and\ \bibinfo {author}
  {\bibfnamefont {K.~F.}\ \bibnamefont {Mak}},\ }\href
  {https://doi.org/10.1038/s41563-019-0294-7} {\bibfield  {journal} {\bibinfo
  {journal} {Nature Materials}\ }\textbf {\bibinfo {volume} {18}},\ \bibinfo
  {pages} {324} (\bibinfo {year} {2019})}\BibitemShut {NoStop}%
\bibitem [{\citenamefont {Du}\ \emph {et~al.}(2021)\citenamefont {Du},
  \citenamefont {Lu},\ and\ \citenamefont {Xie}}]{Du2021Nonlinear}%
  \BibitemOpen
  \bibfield  {author} {\bibinfo {author} {\bibfnamefont {Z.~Z.}\ \bibnamefont
  {Du}}, \bibinfo {author} {\bibfnamefont {H.-Z.}\ \bibnamefont {Lu}},\ and\
  \bibinfo {author} {\bibfnamefont {X.~C.}\ \bibnamefont {Xie}},\ }\href
  {https://doi.org/10.1038/s42254-021-00359-6} {\bibfield  {journal} {\bibinfo
  {journal} {Nature Reviews Physics}\ }\textbf {\bibinfo {volume} {3}},\
  \bibinfo {pages} {744} (\bibinfo {year} {2021})}\BibitemShut {NoStop}%
\bibitem [{\citenamefont {Yu}\ \emph {et~al.}(2014)\citenamefont {Yu},
  \citenamefont {Wu}, \citenamefont {Liu}, \citenamefont {Xu},\ and\
  \citenamefont {Yao}}]{Yu2014Nonlinear}%
  \BibitemOpen
  \bibfield  {author} {\bibinfo {author} {\bibfnamefont {H.}~\bibnamefont
  {Yu}}, \bibinfo {author} {\bibfnamefont {Y.}~\bibnamefont {Wu}}, \bibinfo
  {author} {\bibfnamefont {G.-B.}\ \bibnamefont {Liu}}, \bibinfo {author}
  {\bibfnamefont {X.}~\bibnamefont {Xu}},\ and\ \bibinfo {author}
  {\bibfnamefont {W.}~\bibnamefont {Yao}},\ }\href
  {https://doi.org/10.1103/PhysRevLett.113.156603} {\bibfield  {journal}
  {\bibinfo  {journal} {Phys. Rev. Lett.}\ }\textbf {\bibinfo {volume} {113}},\
  \bibinfo {pages} {156603} (\bibinfo {year} {2014})}\BibitemShut {NoStop}%
\bibitem [{\citenamefont {Rodin}\ \emph {et~al.}(2016)\citenamefont {Rodin},
  \citenamefont {Gomes}, \citenamefont {Carvalho},\ and\ \citenamefont
  {Castro~Neto}}]{Rodin2016Valley}%
  \BibitemOpen
  \bibfield  {author} {\bibinfo {author} {\bibfnamefont {A.~S.}\ \bibnamefont
  {Rodin}}, \bibinfo {author} {\bibfnamefont {L.~C.}\ \bibnamefont {Gomes}},
  \bibinfo {author} {\bibfnamefont {A.}~\bibnamefont {Carvalho}},\ and\
  \bibinfo {author} {\bibfnamefont {A.~H.}\ \bibnamefont {Castro~Neto}},\
  }\href {https://doi.org/10.1103/PhysRevB.93.045431} {\bibfield  {journal}
  {\bibinfo  {journal} {Phys. Rev. B}\ }\textbf {\bibinfo {volume} {93}},\
  \bibinfo {pages} {045431} (\bibinfo {year} {2016})}\BibitemShut {NoStop}%
\bibitem [{\citenamefont {Das}\ \emph {et~al.}(2024)\citenamefont {Das},
  \citenamefont {Ghorai}, \citenamefont {Culcer},\ and\ \citenamefont
  {Agarwal}}]{Das2024Nonlinear}%
  \BibitemOpen
  \bibfield  {author} {\bibinfo {author} {\bibfnamefont {K.}~\bibnamefont
  {Das}}, \bibinfo {author} {\bibfnamefont {K.}~\bibnamefont {Ghorai}},
  \bibinfo {author} {\bibfnamefont {D.}~\bibnamefont {Culcer}},\ and\ \bibinfo
  {author} {\bibfnamefont {A.}~\bibnamefont {Agarwal}},\ }\href
  {https://doi.org/10.1103/PhysRevLett.132.096302} {\bibfield  {journal}
  {\bibinfo  {journal} {Phys. Rev. Lett.}\ }\textbf {\bibinfo {volume} {132}},\
  \bibinfo {pages} {096302} (\bibinfo {year} {2024})}\BibitemShut {NoStop}%
\bibitem [{\citenamefont {Abanin}\ \emph {et~al.}(2009)\citenamefont {Abanin},
  \citenamefont {Shytov}, \citenamefont {Levitov},\ and\ \citenamefont
  {Halperin}}]{Abanin2009Nonlocal}%
  \BibitemOpen
  \bibfield  {author} {\bibinfo {author} {\bibfnamefont {D.~A.}\ \bibnamefont
  {Abanin}}, \bibinfo {author} {\bibfnamefont {A.~V.}\ \bibnamefont {Shytov}},
  \bibinfo {author} {\bibfnamefont {L.~S.}\ \bibnamefont {Levitov}},\ and\
  \bibinfo {author} {\bibfnamefont {B.~I.}\ \bibnamefont {Halperin}},\ }\href
  {https://doi.org/10.1103/PhysRevB.79.035304} {\bibfield  {journal} {\bibinfo
  {journal} {Phys. Rev. B}\ }\textbf {\bibinfo {volume} {79}},\ \bibinfo
  {pages} {035304} (\bibinfo {year} {2009})}\BibitemShut {NoStop}%
\bibitem [{\citenamefont {Beconcini}\ \emph {et~al.}(2016)\citenamefont
  {Beconcini}, \citenamefont {Taddei},\ and\ \citenamefont
  {Polini}}]{Beconcini2016Nonlocal}%
  \BibitemOpen
  \bibfield  {author} {\bibinfo {author} {\bibfnamefont {M.}~\bibnamefont
  {Beconcini}}, \bibinfo {author} {\bibfnamefont {F.}~\bibnamefont {Taddei}},\
  and\ \bibinfo {author} {\bibfnamefont {M.}~\bibnamefont {Polini}},\ }\href
  {https://doi.org/10.1103/PhysRevB.94.121408} {\bibfield  {journal} {\bibinfo
  {journal} {Phys. Rev. B}\ }\textbf {\bibinfo {volume} {94}},\ \bibinfo
  {pages} {121408} (\bibinfo {year} {2016})}\BibitemShut {NoStop}%
\bibitem [{\citenamefont {Gao}\ \emph {et~al.}(2023)\citenamefont {Gao},
  \citenamefont {Liu}, \citenamefont {Qiu}, \citenamefont {Ghosh},
  \citenamefont {V.~Trevisan}, \citenamefont {Onishi}, \citenamefont {Hu},
  \citenamefont {Qian}, \citenamefont {Tien}, \citenamefont {Chen},
  \citenamefont {Huang}, \citenamefont {Bérubé}, \citenamefont {Li},
  \citenamefont {Tzschaschel}, \citenamefont {Dinh}, \citenamefont {Sun},
  \citenamefont {Ho}, \citenamefont {Lien}, \citenamefont {Singh},
  \citenamefont {Watanabe}, \citenamefont {Taniguchi}, \citenamefont {Bell},
  \citenamefont {Lin}, \citenamefont {Chang}, \citenamefont {Du}, \citenamefont
  {Bansil}, \citenamefont {Fu}, \citenamefont {Ni}, \citenamefont {Orth},
  \citenamefont {Ma},\ and\ \citenamefont {Xu}}]{Gao2023QM}%
  \BibitemOpen
  \bibfield  {author} {\bibinfo {author} {\bibfnamefont {A.}~\bibnamefont
  {Gao}}, \bibinfo {author} {\bibfnamefont {Y.-F.}\ \bibnamefont {Liu}},
  \bibinfo {author} {\bibfnamefont {J.-X.}\ \bibnamefont {Qiu}}, \bibinfo
  {author} {\bibfnamefont {B.}~\bibnamefont {Ghosh}}, \bibinfo {author}
  {\bibfnamefont {T.}~\bibnamefont {V.~Trevisan}}, \bibinfo {author}
  {\bibfnamefont {Y.}~\bibnamefont {Onishi}}, \bibinfo {author} {\bibfnamefont
  {C.}~\bibnamefont {Hu}}, \bibinfo {author} {\bibfnamefont {T.}~\bibnamefont
  {Qian}}, \bibinfo {author} {\bibfnamefont {H.-J.}\ \bibnamefont {Tien}},
  \bibinfo {author} {\bibfnamefont {S.-W.}\ \bibnamefont {Chen}}, \bibinfo
  {author} {\bibfnamefont {M.}~\bibnamefont {Huang}}, \bibinfo {author}
  {\bibfnamefont {D.}~\bibnamefont {Bérubé}}, \bibinfo {author}
  {\bibfnamefont {H.}~\bibnamefont {Li}}, \bibinfo {author} {\bibfnamefont
  {C.}~\bibnamefont {Tzschaschel}}, \bibinfo {author} {\bibfnamefont
  {T.}~\bibnamefont {Dinh}}, \bibinfo {author} {\bibfnamefont {Z.}~\bibnamefont
  {Sun}}, \bibinfo {author} {\bibfnamefont {S.-C.}\ \bibnamefont {Ho}},
  \bibinfo {author} {\bibfnamefont {S.-W.}\ \bibnamefont {Lien}}, \bibinfo
  {author} {\bibfnamefont {B.}~\bibnamefont {Singh}}, \bibinfo {author}
  {\bibfnamefont {K.}~\bibnamefont {Watanabe}}, \bibinfo {author}
  {\bibfnamefont {T.}~\bibnamefont {Taniguchi}}, \bibinfo {author}
  {\bibfnamefont {D.~C.}\ \bibnamefont {Bell}}, \bibinfo {author}
  {\bibfnamefont {H.}~\bibnamefont {Lin}}, \bibinfo {author} {\bibfnamefont
  {T.-R.}\ \bibnamefont {Chang}}, \bibinfo {author} {\bibfnamefont {C.~R.}\
  \bibnamefont {Du}}, \bibinfo {author} {\bibfnamefont {A.}~\bibnamefont
  {Bansil}}, \bibinfo {author} {\bibfnamefont {L.}~\bibnamefont {Fu}}, \bibinfo
  {author} {\bibfnamefont {N.}~\bibnamefont {Ni}}, \bibinfo {author}
  {\bibfnamefont {P.~P.}\ \bibnamefont {Orth}}, \bibinfo {author}
  {\bibfnamefont {Q.}~\bibnamefont {Ma}},\ and\ \bibinfo {author}
  {\bibfnamefont {S.-Y.}\ \bibnamefont {Xu}},\ }\href
  {https://doi.org/10.1126/science.adf1506} {\bibfield  {journal} {\bibinfo
  {journal} {Science}\ }\textbf {\bibinfo {volume} {381}},\ \bibinfo {pages}
  {181} (\bibinfo {year} {2023})}\BibitemShut {NoStop}%
\bibitem [{\citenamefont {Wang}\ \emph {et~al.}(2023)\citenamefont {Wang},
  \citenamefont {Kaplan}, \citenamefont {Zhang}, \citenamefont {Holder},
  \citenamefont {Cao}, \citenamefont {Wang}, \citenamefont {Zhou},
  \citenamefont {Zhou}, \citenamefont {Jiang}, \citenamefont {Zhang},
  \citenamefont {Ru}, \citenamefont {Cai}, \citenamefont {Watanabe},
  \citenamefont {Taniguchi}, \citenamefont {Yan},\ and\ \citenamefont
  {Gao}}]{Wang2023QM}%
  \BibitemOpen
  \bibfield  {author} {\bibinfo {author} {\bibfnamefont {N.}~\bibnamefont
  {Wang}}, \bibinfo {author} {\bibfnamefont {D.}~\bibnamefont {Kaplan}},
  \bibinfo {author} {\bibfnamefont {Z.}~\bibnamefont {Zhang}}, \bibinfo
  {author} {\bibfnamefont {T.}~\bibnamefont {Holder}}, \bibinfo {author}
  {\bibfnamefont {N.}~\bibnamefont {Cao}}, \bibinfo {author} {\bibfnamefont
  {A.}~\bibnamefont {Wang}}, \bibinfo {author} {\bibfnamefont {X.}~\bibnamefont
  {Zhou}}, \bibinfo {author} {\bibfnamefont {F.}~\bibnamefont {Zhou}}, \bibinfo
  {author} {\bibfnamefont {Z.}~\bibnamefont {Jiang}}, \bibinfo {author}
  {\bibfnamefont {C.}~\bibnamefont {Zhang}}, \bibinfo {author} {\bibfnamefont
  {S.}~\bibnamefont {Ru}}, \bibinfo {author} {\bibfnamefont {H.}~\bibnamefont
  {Cai}}, \bibinfo {author} {\bibfnamefont {K.}~\bibnamefont {Watanabe}},
  \bibinfo {author} {\bibfnamefont {T.}~\bibnamefont {Taniguchi}}, \bibinfo
  {author} {\bibfnamefont {B.}~\bibnamefont {Yan}},\ and\ \bibinfo {author}
  {\bibfnamefont {W.}~\bibnamefont {Gao}},\ }\href
  {https://doi.org/10.1038/s41586-023-06363-3} {\bibfield  {journal} {\bibinfo
  {journal} {Nature}\ }\textbf {\bibinfo {volume} {621}},\ \bibinfo {pages}
  {487} (\bibinfo {year} {2023})}\BibitemShut {NoStop}%
\bibitem [{\citenamefont {Han}\ \emph {et~al.}(2024)\citenamefont {Han},
  \citenamefont {Uchimura}, \citenamefont {Araki}, \citenamefont {Yoon},
  \citenamefont {Takeuchi}, \citenamefont {Yamane}, \citenamefont {Kanai},
  \citenamefont {Ieda}, \citenamefont {Ohno},\ and\ \citenamefont
  {Fukami}}]{Han2024room}%
  \BibitemOpen
  \bibfield  {author} {\bibinfo {author} {\bibfnamefont {J.}~\bibnamefont
  {Han}}, \bibinfo {author} {\bibfnamefont {T.}~\bibnamefont {Uchimura}},
  \bibinfo {author} {\bibfnamefont {Y.}~\bibnamefont {Araki}}, \bibinfo
  {author} {\bibfnamefont {J.-Y.}\ \bibnamefont {Yoon}}, \bibinfo {author}
  {\bibfnamefont {Y.}~\bibnamefont {Takeuchi}}, \bibinfo {author}
  {\bibfnamefont {Y.}~\bibnamefont {Yamane}}, \bibinfo {author} {\bibfnamefont
  {S.}~\bibnamefont {Kanai}}, \bibinfo {author} {\bibfnamefont
  {J.}~\bibnamefont {Ieda}}, \bibinfo {author} {\bibfnamefont {H.}~\bibnamefont
  {Ohno}},\ and\ \bibinfo {author} {\bibfnamefont {S.}~\bibnamefont {Fukami}},\
  }\href {https://doi.org/10.1038/s41567-024-02476-2} {\bibfield  {journal}
  {\bibinfo  {journal} {Nat. Phys.}\ }\textbf {\bibinfo {volume} {20}},\
  \bibinfo {pages} {1110} (\bibinfo {year} {2024})}\BibitemShut {NoStop}%
\bibitem [{\citenamefont {Wang}\ \emph {et~al.}(2021)\citenamefont {Wang},
  \citenamefont {Gao},\ and\ \citenamefont {Xiao}}]{wang2021}%
  \BibitemOpen
  \bibfield  {author} {\bibinfo {author} {\bibfnamefont {C.}~\bibnamefont
  {Wang}}, \bibinfo {author} {\bibfnamefont {Y.}~\bibnamefont {Gao}},\ and\
  \bibinfo {author} {\bibfnamefont {D.}~\bibnamefont {Xiao}},\ }\href
  {https://doi.org/10.1103/PhysRevLett.127.277201} {\bibfield  {journal}
  {\bibinfo  {journal} {Phys. Rev. Lett.}\ }\textbf {\bibinfo {volume} {127}},\
  \bibinfo {pages} {277201} (\bibinfo {year} {2021})}\BibitemShut {NoStop}%
\bibitem [{\citenamefont {Liu}\ \emph {et~al.}(2021)\citenamefont {Liu},
  \citenamefont {Zhao}, \citenamefont {Huang}, \citenamefont {Wu},
  \citenamefont {Sheng}, \citenamefont {Xiao},\ and\ \citenamefont
  {Yang}}]{liu2021}%
  \BibitemOpen
  \bibfield  {author} {\bibinfo {author} {\bibfnamefont {H.}~\bibnamefont
  {Liu}}, \bibinfo {author} {\bibfnamefont {J.}~\bibnamefont {Zhao}}, \bibinfo
  {author} {\bibfnamefont {Y.-X.}\ \bibnamefont {Huang}}, \bibinfo {author}
  {\bibfnamefont {W.}~\bibnamefont {Wu}}, \bibinfo {author} {\bibfnamefont
  {X.-L.}\ \bibnamefont {Sheng}}, \bibinfo {author} {\bibfnamefont
  {C.}~\bibnamefont {Xiao}},\ and\ \bibinfo {author} {\bibfnamefont {S.~A.}\
  \bibnamefont {Yang}},\ }\href
  {https://doi.org/10.1103/PhysRevLett.127.277202} {\bibfield  {journal}
  {\bibinfo  {journal} {Phys. Rev. Lett.}\ }\textbf {\bibinfo {volume} {127}},\
  \bibinfo {pages} {277202} (\bibinfo {year} {2021})}\BibitemShut {NoStop}%
\bibitem [{\citenamefont {Du}\ \emph {et~al.}(2019)\citenamefont {Du},
  \citenamefont {Wang}, \citenamefont {Li}, \citenamefont {Lu},\ and\
  \citenamefont {Xie}}]{Du2019Disorder}%
  \BibitemOpen
  \bibfield  {author} {\bibinfo {author} {\bibfnamefont {Z.~Z.}\ \bibnamefont
  {Du}}, \bibinfo {author} {\bibfnamefont {C.~M.}\ \bibnamefont {Wang}},
  \bibinfo {author} {\bibfnamefont {S.}~\bibnamefont {Li}}, \bibinfo {author}
  {\bibfnamefont {H.-Z.}\ \bibnamefont {Lu}},\ and\ \bibinfo {author}
  {\bibfnamefont {X.~C.}\ \bibnamefont {Xie}},\ }\href
  {https://doi.org/10.1038/s41467-019-10941-3} {\bibfield  {journal} {\bibinfo
  {journal} {Nature Communications}\ }\textbf {\bibinfo {volume} {10}},\
  \bibinfo {pages} {3047} (\bibinfo {year} {2019})}\BibitemShut {NoStop}%
\bibitem [{\citenamefont {Huang}\ \emph {et~al.}()\citenamefont {Huang},
  \citenamefont {Xiao}, \citenamefont {Yang},\ and\ \citenamefont
  {Li}}]{Huang2023Scaling}%
  \BibitemOpen
  \bibfield  {author} {\bibinfo {author} {\bibfnamefont {Y.-X.}\ \bibnamefont
  {Huang}}, \bibinfo {author} {\bibfnamefont {C.}~\bibnamefont {Xiao}},
  \bibinfo {author} {\bibfnamefont {S.~A.}\ \bibnamefont {Yang}},\ and\
  \bibinfo {author} {\bibfnamefont {X.}~\bibnamefont {Li}},\ }\href@noop {}
  {}\Eprint {https://arxiv.org/abs/2311.01219} {arXiv:2311.01219} \BibitemShut
  {NoStop}%
\bibitem [{\citenamefont {Ihn}(2009)}]{Ihn2009Semiconductor}%
  \BibitemOpen
  \bibfield  {author} {\bibinfo {author} {\bibfnamefont {T.}~\bibnamefont
  {Ihn}},\ }\href@noop {} {\emph {\bibinfo {title} {Semiconductor
  Nanostructures: Quantum states and electronic transport}}}\ (\bibinfo
  {publisher} {Oxford University Press, Oxford, New York},\ \bibinfo {year}
  {2009})\BibitemShut {NoStop}%
\bibitem [{sup()}]{supp}%
  \BibitemOpen
  \href@noop {} {\bibinfo  {journal} {See the Supplemental Material for details
  of nonlocal nonlinear transport theory and of symmetry analysis, and the
  computational methods}\ }\BibitemShut {NoStop}%
\bibitem [{\citenamefont {Lai}\ \emph {et~al.}(2021)\citenamefont {Lai},
  \citenamefont {Liu}, \citenamefont {Zhang}, \citenamefont {Zhao},
  \citenamefont {Feng}, \citenamefont {Wang}, \citenamefont {Tang},
  \citenamefont {Liu}, \citenamefont {Novoselov}, \citenamefont {Yang},\ and\
  \citenamefont {Gao}}]{Lai2021}%
  \BibitemOpen
\bibfield  {journal} {  }\bibfield  {author} {\bibinfo {author} {\bibfnamefont
  {S.}~\bibnamefont {Lai}}, \bibinfo {author} {\bibfnamefont {H.}~\bibnamefont
  {Liu}}, \bibinfo {author} {\bibfnamefont {Z.}~\bibnamefont {Zhang}}, \bibinfo
  {author} {\bibfnamefont {J.}~\bibnamefont {Zhao}}, \bibinfo {author}
  {\bibfnamefont {X.}~\bibnamefont {Feng}}, \bibinfo {author} {\bibfnamefont
  {N.}~\bibnamefont {Wang}}, \bibinfo {author} {\bibfnamefont {C.}~\bibnamefont
  {Tang}}, \bibinfo {author} {\bibfnamefont {Y.}~\bibnamefont {Liu}}, \bibinfo
  {author} {\bibfnamefont {K.~S.}\ \bibnamefont {Novoselov}}, \bibinfo {author}
  {\bibfnamefont {S.~A.}\ \bibnamefont {Yang}},\ and\ \bibinfo {author}
  {\bibfnamefont {W.-b.}\ \bibnamefont {Gao}},\ }\href
  {https://doi.org/10.1038/s41565-021-00917-0} {\bibfield  {journal} {\bibinfo
  {journal} {Nature Nanotechnology}\ }\textbf {\bibinfo {volume} {16}},\
  \bibinfo {pages} {869} (\bibinfo {year} {2021})}\BibitemShut {NoStop}%
\bibitem [{\citenamefont {Zheng}\ \emph {et~al.}(2016)\citenamefont {Zheng},
  \citenamefont {Cai}, \citenamefont {Ge}, \citenamefont {Zhang}, \citenamefont
  {Liu}, \citenamefont {Lu}, \citenamefont {Zhang}, \citenamefont {Qiu},
  \citenamefont {Taniguchi}, \citenamefont {Watanabe}, \citenamefont {Jia},
  \citenamefont {Qi}, \citenamefont {Chen}, \citenamefont {Sun},\ and\
  \citenamefont {Feng}}]{Zheng2016Quantum}%
  \BibitemOpen
  \bibfield  {author} {\bibinfo {author} {\bibfnamefont {F.}~\bibnamefont
  {Zheng}}, \bibinfo {author} {\bibfnamefont {C.}~\bibnamefont {Cai}}, \bibinfo
  {author} {\bibfnamefont {S.}~\bibnamefont {Ge}}, \bibinfo {author}
  {\bibfnamefont {X.}~\bibnamefont {Zhang}}, \bibinfo {author} {\bibfnamefont
  {X.}~\bibnamefont {Liu}}, \bibinfo {author} {\bibfnamefont {H.}~\bibnamefont
  {Lu}}, \bibinfo {author} {\bibfnamefont {Y.}~\bibnamefont {Zhang}}, \bibinfo
  {author} {\bibfnamefont {J.}~\bibnamefont {Qiu}}, \bibinfo {author}
  {\bibfnamefont {T.}~\bibnamefont {Taniguchi}}, \bibinfo {author}
  {\bibfnamefont {K.}~\bibnamefont {Watanabe}}, \bibinfo {author}
  {\bibfnamefont {S.}~\bibnamefont {Jia}}, \bibinfo {author} {\bibfnamefont
  {J.}~\bibnamefont {Qi}}, \bibinfo {author} {\bibfnamefont {J.-H.}\
  \bibnamefont {Chen}}, \bibinfo {author} {\bibfnamefont {D.}~\bibnamefont
  {Sun}},\ and\ \bibinfo {author} {\bibfnamefont {J.}~\bibnamefont {Feng}},\
  }\href {https://doi.org/https://doi.org/10.1002/adma.201600100} {\bibfield
  {journal} {\bibinfo  {journal} {Advanced Materials}\ }\textbf {\bibinfo
  {volume} {28}},\ \bibinfo {pages} {4845} (\bibinfo {year}
  {2016})}\BibitemShut {NoStop}%
\end{thebibliography}%

\end{document}